\PassOptionsToPackage{dvipsnames}{xcolor} 
\documentclass[twocolumn]{aastex631}

\usepackage{enumitem}
\usepackage{graphicx}	          
\usepackage{amsmath}	          
\usepackage{upgreek}            
\usepackage{amssymb}	          
\usepackage{bm}		              
\usepackage{CJKutf8}            
\usepackage[T1]{fontenc}        
\usepackage{comment}            
\usepackage{tabularx}           
\usepackage{threeparttable}     
\usepackage{booktabs}           
\usepackage{soul}               
\usepackage{ulem}               
\usepackage{multirow}           
\usepackage{microtype}          
\DisableLigatures[f]{encoding = *, family = *} 
\usepackage[figuresleft]{rotating}  
\allowdisplaybreaks             
\usepackage{afterpage}          
\emergencystretch=\maxdimen     
\def\numx#1e#2{{#1}\mathrm{e}{#2}}

\newcommand{\fracp}[2]{\frac{\partial{#1}}{\partial{#2}}}

\newcommand{\mwSt}{\langle {\rm St} \rangle}
\newcommand{\Kdelta}{\boldsymbol{\updelta}}

\newcommand{\rlr}[1]{#1} 

\makeatletter

\newenvironment{newfigure*}
  {\renewcommand{\fnum@figure}{\textcolor{blue}{\bfseries Figure~\thefigure}}%
   \begin{figure*}}
  {\end{figure*}}

\newenvironment{revfigure*}
  {\renewcommand{\fnum@figure}{\textcolor{blue}{\bfseries Figure~\thefigure}}%
   \begin{figure*}}
  {\end{figure*}}
\makeatother

\defcitealias{Kunitomo2020}{KSI20}

\received{--}
\revised{--}
\accepted{--}
\submitjournal{ApJ}

\shorttitle{CCKB Backstory}
\shortauthors{Li and Chiang}

\begin{document}

\title{The Edges of Planetary Systems: Falling Off the Kuiper Cliff in a Dissipating Gas Disk}

\correspondingauthor{Rixin Li}
\email{rixin@berkeley.edu}
\author[0000-0001-9222-4367]{Rixin Li
\begin{CJK*}{UTF8}{gkai}(李日新)\end{CJK*}
}
\altaffiliation{51 Pegasi b Fellow}
\affiliation{Department of Astronomy, Theoretical Astrophysics Center, and Center for Integrative Planetary Science, University of California Berkeley, Berkeley, CA 94720-3411, USA}

\author[0000-0002-6246-2310]{Eugene Chiang
\begin{CJK*}{UTF8}{bkai}(蔣詒曾)\end{CJK*}}
\affiliation{Department of Astronomy, Theoretical Astrophysics Center, and Center for Integrative Planetary Science, University of California Berkeley, Berkeley, CA 94720-3411, USA}
\affiliation{Department of Earth and Planetary Science, University of California Berkeley, Berkeley, CA 94720-4767, USA}

\begin{abstract}
Probably the last planetesimals to have formed from dust in the solar nebula are Cold Classical Kuiper belt objects (CCKBOs). To the extent that they are isolated and unchanged since birth, CCKBOs offer direct insights into nebular processes. Their population density drops abruptly beyond a heliocentric radius of $\sim$47 au, a feature known as the ``Kuiper Cliff''. We show with global, 1D (radial), time-dependent models how gaseous protoplanetary disks that disperse from magnetic and photoevaporative winds leave behind planetesimal disks with Cliff-like outer edges. The gas disperses from the inside out, creating transitional disks whose inner cavities expand from $\lesssim$ 1 au to $\gtrsim$ 100 au. Gas at the cavity boundary presents a pressure maximum toward which dust particles drift, triggering the streaming instability which clumps dust into planetesimals massive enough to decouple from gas. The receding cavity wall thus paves a disk of planetesimals which truncates when dust and gas are spent. With no fine-tuning, we show how a generic gas disk clearing from the inside out reproduces the Kuiper Cliff and the CCKB surface density. Connecting these global 1D results with published local 3D simulations of dust and gas, we see how many properties of the CCKB --- its radial extent, total mass, individual object sizes, and binary statistics --- follow from the streaming instability at work in a late-stage transition disk.
\end{abstract}


\section{Introduction}
\label{sec:intro}

For most of its history, the subject of planetesimal formation has been consigned to theory, as observations of protoplanetary disks are largely incapable of detecting bodies intermediate in size between dust grains and planets. While objects bridging this gap have been deduced in extrasolar debris disks, the inferred bodies are still not directly visible and have properties not uniquely constrained (e.g.~\citealt{Jankovic2026}; \citealt{Chiang2026}); in any case, over a wide range of parameter space they are grinding down to dust, not building up. The same wrong sign applies to the Solar System's old and weathered main asteroid belt. The Kuiper belt is today collisionless, but tracing its formation back to the primordial circumsolar disk is made difficult by most Kuiper belt objects having been gravitationally scattered far from their birthplaces (\citealt{Morbidelli2020}; \citealt{Gladman_Volk2021}).

Arguably the best if not the only place we can look for observational clues to  planetesimal formation --- including preserved first-generation planetesimals --- is in the trans-Neptunian sub-population of Cold Classical Kuiper belt objects (CCKBOs).  The largest of these, having individual radii of $\sim$100 km and frequently found in delicate, wide-separation, equal-size binaries (e.g.~\citealt{Noll2020}), execute heliocentric orbits of low eccentricity and inclination ($e,i \lesssim 0.1$) between 42 and 47 au from the Sun (e.g.~\citealt{Brown2001}; \citealt{Van_Laerhoven2019}; \citealt{Batygin2020}; \citealt{Kavelaars2021}; \citealt{Huang2022}; \citealt{Fraser2024}).  With their distinctively colored surfaces (e.g.~\citealt{Gulbis2006}; \citealt{Fraser2017}; \citealt{Fraser2021}), CCKBOs are thought to have coalesced directly from the solar nebula and to have resided at their current orbital distances their entire lives, relatively untouched by collisions and largely shielded from the chaos that produced the Oort Cloud and dynamically hot components of the Kuiper belt (e.g.~\citealt{Nesvorny2019b}; \citealt{Nesvorny2022}).

Only a few thousandths of an Earth mass (give or take a factor of a few) is contained in the CCKB \citep{Petit2023,Napier2024}; the corresponding number of 100-km-radius objects is a few thousand. As this mass is tiny compared to the solid (non-H, non-He) mass in the rest of the Solar System, and as the CCKB is observed to cut off at its outer edge at $\sim$47 au (around the so-called ``Kuiper Cliff''; see e.g.~Fig.~4 of \citealt{Fraser2024}), it is natural to suppose that the CCKB contains the most radially distant and chronologically last planetesimals to have formed from a dissipating solar nebula. In support of the idea that the CCKB coagulated from the dregs of the protoplanetary disk, \citet{Li_Chiang_2025} showed how some basic properties of CCKBOs --- their individual masses, total numbers, and the incidence of prograde and retrograde binaries (see also \citealt{Nesvorny2019}) --- could be reproduced in a disk having a gas surface density of $\sim$0.1--0.4 g/cm$^2$ at $\sim$45 au, amounting to just a few percent of the local surface density in the primordial minimum-mass solar nebula.  According to the calculations of \citet{Li_Chiang_2025}, the CCKB forms when dust particles trigger the streaming instability (SI; \citealt{Youdin_Goodman_2005}) between 42--47 au as they drift radially through this region. The SI-induced turbulence clumps the dust, and the strongest overdensities collapse from self-gravity into bound, fast-spinning proto-CCKBOs that are massive enough to decouple from the background particle drift.

\citet{Li_Chiang_2025} leave unspecified how their ``drift-and-form'' scenario is set up in the first place. They conduct local shearing box simulations of dust particles that start in the CCKB volume and drift through, but do not address larger-scale issues: what happens outside the CCKB volume and at earlier times; what justifies the dust aerodynamic stopping times which they assumed to be comparable to the local orbital time, maximizing SI growth rates; and most crucially, what formed the Kuiper Cliff, i.e.~why planetesimals do not also form at larger orbital distances.

The goal of this paper is to answer these questions with a bigger picture; we seek to provide a backstory for the background particle drift. We construct a global, 1D (in stellocentric radius $r$), time-dependent model for how dust evolves in a gaseous protoplanetary disk, starting from gas-rich conditions, and ending with the nebula's complete dissipation. The gas  evolves following current thinking about transport by magneto-hydrodynamic disk winds (\citealt{Bai2013a}; \citealt{Bai2016}, and references therein) and mass loss by photoevaporation from stellar X-ray and ultraviolet radiation (e.g.~\citealt{Clarke2001}; \citealt{Alexander2006a, Alexander2006b}; \citealt{Gorti2009}; \citealt{Owen2010}; \citealt{Alexander2014}; \citealt{Kunitomo2021}). We utilize the 1D gas disk models of \citet{Kunitomo2020} which incorporate these gas transport processes across multiple decades in $r$ and over several Myrs. Grafted onto the gas evolution is the dust evolution, which we model using  \texttt{DustPy} for the transport, growth, and fragmentation of dust \citep{dustpy}, plus a prescription for planetesimal formation drawn from the SI simulations of \citet{Li_Chiang_2025}. The computed surface density profile of planetesimals is then compared with the observed present-day surface density of the CCKB: $\Sigma_{\rm CCKB} \sim 10^{-4}$ g/cm$^2$ at $r \simeq 40$--50 au. While we do not expect to match these numbers exactly  as the \citetalias{Kunitomo2020} disk models are generic and contain zero prior knowledge of the CCKB, we do hope to learn something about how the outermost edges of planetary systems, as defined by their first-generation planetesimal disks, are shaped. As it turns out, we will achieve remarkably good quantitative agreement with the CCKB.

Section \ref{sec:method} contains our methods, and Section \ref{sec:results}  results. We summarize and provide an outlook in Section \ref{sec:summary}.

\section{Method}
\label{sec:method}

Our method in outline is to take the time-dependent gas disk models of \citeauthor{Kunitomo2020}~(\citeyear{Kunitomo2020}, hereafter \citetalias{Kunitomo2020}), and solve for the transport and evolution of dust within these gas models. Thus we do not ourselves calculate how the gas evolves, but take as given the gas surface density $\Sigma_{\rm g}$ as a function of stellocentric radius $r$ and time $t$. The dust surface density evolution $\Sigma_{\rm d}(r,t)$, and by extension the formation of planetesimals as traced by their surface density $\Sigma_{\rm pl}(r,t)$, are then calculated by inserting $\Sigma_{\rm g}(r,t)$ from \citetalias{Kunitomo2020} into \texttt{DustPy} \citep{dustpy}, which models the 1D radial transport of dust by gas advection and diffusion, and the evolution of the dust size distribution from coagulation and fragmentation. For this paper we have developed a new GPU-accelerated port of \texttt{DustPy}, which we call \texttt{DustPy-GPU}
\footnote{Repository: \href{https://github.com/astroboylrx/dustpy-gpu}{https://github.com/astroboylrx/dustpy-gpu}}. 

Our procedure is not strictly self-consistent because it utilizes the \citetalias{Kunitomo2020} gas disk models which make assumptions (typically inherited from still earlier work on which the models are based) about disk solids and gas-phase metallicities that are not necessarily satisfied by our \texttt{DustPy-GPU} models. We assume the inconsistencies are ignorable, and have reasons to believe they are. The magnetized disk winds and photoevaporative winds that drive much of the \citetalias{Kunitomo2020} gas evolution are confined to externally ionized disk surface layers containing only a tiny fraction of the total supply of metals (e.g.~\citealt{Perez-Becker2011}), whereas \texttt{DustPy-GPU} evolves the bulk of the solids located near the midplane. Even at the midplane, changes to the dust size distribution have little effect on the temperature and gas scale height (e.g.~\citealt{Chiang2001}).

\begingroup 
\setlength{\medmuskip}{0mu} 
\begin{deluxetable*}{cccccccccc}[ht]
  \tablecaption{Simulation parameters and outcomes}\label{tab:paras}
  \tablecolumns{10}
  \tablehead{
    \colhead{(1)} &
    \colhead{(2)} &
    \colhead{(3)} &
    \colhead{(4)} &
    \colhead{(5)} &
    \colhead{(6)} &
    \colhead{(7)} &
    \colhead{(8)} &
    \colhead{(9)} & 
    \colhead{(10)} \\
    \colhead{\texttt{Run}} &
    \colhead{KSI20 gas model} &
    \colhead{$\alpha$} &
    \colhead{$\delta$} &
    \colhead{$Z$} &    
    \colhead{$M_{\rm dust,init}$} &
    \colhead{$M_{\rm dust,final}$} &
    \colhead{$M_{\rm pl,total}$} &
    \colhead{$M_{\rm pl,main}$} &
    \colhead{$r_{\rm main,out}$} \\
    \colhead{} &
    \colhead{} &
    \colhead{} &
    \colhead{} &
    \colhead{} &
    \colhead{[$M_{\oplus}$]} &
    \colhead{[$M_{\oplus}$]} &
    \colhead{[$M_{\oplus}$]} &
    \colhead{[$M_{\oplus}$]} &
    \colhead{[au]}   
  }
  \startdata
  \hline
  \texttt{Fid}     &\texttt{C-i} &$0$ &$\delta(\mwSt)$  &$0.01$  &$317.07$ &$0.12$  &$74.86$ &$46.05$ &$41.9$--$49.2$ \\ 
  \texttt{Fid-hiZ} &\texttt{C-i} &$0$ &$\delta(\mwSt)$  &$0.03$  &$951.21$ &$475.0$ &$230.9$ &$1.22$  &$41.9$--$49.2$ \\
  \texttt{Fid-loZ} &\texttt{C-i} &$0$ &$\delta(\mwSt)$  &$0.003$ &$95.12$  &$0.05$  &$0.62$  &$0.41$  &$42.6$--$49.2$ \\
  \texttt{Fid-$\Pi$} &\texttt{C-i} &$0$ &$\delta(\mwSt)$  &$0.01$  &$317.07$ &$24.17$ &$25.86$ &$18.71$  &$45.8$--$52.8$ \\
  \texttt{Fid-$\Sigma$} &\texttt{C-i} &$0$ &$\delta(\mwSt)$  &$0.01$  &$317.07$ &$0.11$ &$74.86$ &$46.05$  &$41.9$--$54.7$ \\
  \texttt{PEW}     &\texttt{A-i} &$8\times10^{-5}$ &$8\times10^{-5}$  &$0.01$  &$317.07$ &$0.02$  &$0.03$ &$0.03$ & $14.9$ \\
  \enddata
  \tablecomments{Columns: 
  (1) Run name;
  (2) Gas model in \citetalias{Kunitomo2020}, where \texttt{C-i} includes MDW and PEW, and \texttt{A-i} includes only PEW (Section \ref{subsec:disk_models});
  (3) Dimensionless gas diffusivity used to calculate $v_{\rm visc}$, the radial velocity of gas due to viscous diffusion (eqn.~\ref{eq:vvisc});
  (4) Dimensionless dust diffusivity  (eqn.~\ref{eq:dust_diffusion} and surrounding text), which in our \texttt{Fid} models depends on $\langle {\rm St} \rangle$, the mass-weighted Stokes number of dust particles;
  (5) Initial global dust-to-gas ratio;
  (6) Initial total dust mass (dust particles are initially micron-sized);
  (7) Final total dust mass remaining in the simulation domain at the end of the run;
  (8) Final total mass in planetesimals formed;
  (9) Final mass in planetesimals in the ``main disk'' located away from simulation boundaries;
  (10) Range of radii spanned by the outer edge of the main planetesimal disk, starting from where $\Sigma_{\rm pl}=\Sigma_{\rm CCKB} = 10^{-4}$ g/cm$^{2}$ and ending where $\Sigma_{\rm pl}$ vanishes (see e.g.~Figs.~\ref{fig:run_Fid}--\ref{fig:run_Fid-varZ-final}). For \texttt{Run PEW}, planetesimals are formed within only one radial grid cell.
  \rlr{Simulation videos are available at \href{https://www.rixinli.com/cckb}{rixinli.com/CCKB}.}
  }
\end{deluxetable*}
\endgroup

\subsection{Gas disk models}
\label{subsec:disk_models}

\citetalias{Kunitomo2020} calculate how a protoplanetary disk orbiting a $1M_\odot$ star  accretes and disperses by the action of a  magnetized disk wind (MDW) and/or a thermally driven photoevaporative wind (PEW). For our fiducial test case (\texttt{Fid}, Table \ref{tab:paras}) we adopt \texttt{Model C-i} of \citetalias{Kunitomo2020} which \rlr{includes both \texttt{PEW} and a strong \texttt{MDW} (their Figure 2), and which they prefer as the model most supported by observations (e.g.~\citealt{Mamajek2009,Flaherty2017,Ercolano2017,Pascucci2025}).} We also experiment with their \texttt{Model A-i} (their Figure 1) which accounts for \texttt{PEW} only.  In each model, the initial gas disk mass is $0.118 M_\odot$. 

The gas surface densities $\Sigma_{\rm g}(r,t)$, gas temperatures $T(r,t)$, and gas scale heights $H_{\rm g}(r,t)$ are tabulated by \citetalias{Kunitomo2020} across $2000$ logarithmically spaced grid points from $r = 10^{-2}$ to $10^4$ au, and over 307 snapshots in time from $t=0$ to $t = 6.22$ Myr for \texttt{Model C-i}, and 95 snapshots from $t=0$ to $t = 15.85$ Myr for \texttt{Model A-i}.  Whatever gas quantities are required by our dust solver are obtained by linear interpolation (\texttt{scipy.interpolate.RegularGridInterpolator}) in radius and time over our own grids (Section \ref{subsec:BCs}).

\subsection{Governing equations for dust}
\label{subsec:eqs}

At $t=0$, the dust surface density $\Sigma_{\rm d}(r) = Z \Sigma_{\rm g}(r)$ with constant dust-to-gas ratio $Z$. Our fiducial $Z = 0.01$, and we experiment with high $Z=0.03$ (\texttt{Fid-hiZ}, Table \ref{tab:paras}) and low $Z = 0.003$ (\texttt{Fid-loZ}), \rlr{values motivated by disk surveys (e.g., \citealt{Ansdell2016}; \citealt{Trapman2025})}. The initial dust size distribution is an MRN \citep{MRN} distribution running from a grain radius of 0.5 to 1 $\mu$m.

The equations and algorithms underlying  \texttt{DustPy-GPU} are in \citet{dustpy} and \citet{Birnstiel2010}. Here we provide a summary. At every $r$, the dust surface density evolves according to the continuity equation:
\begin{align}
  \fracp{\Sigma_{{\rm d},i}}{t} + \frac{1}{r}\fracp{}{r}\left[r\Sigma_{{\rm d},i}v_{{\rm d},i} - rD_{i}\Sigma_{\rm g}\fracp{}{r}\left(\frac{\Sigma_{{\rm d},i}}{\Sigma_{\rm g}}\right)\right] = S_{{\rm ext},i} \label{eq:sigma_d}
  \end{align}
where $i$ indexes on dust size. The first term in square brackets accounts for laminar drift of dust due to gas drag at velocity
\begin{align}
  v_{{\rm d},i} &= \left(v_{\rm g} + 2 v_{\rm drift}^{\rm max} {\rm St}_i\right) \frac{1}{{\rm St}_i^2 + 1}
  \end{align}
where
\begin{align}
  v_{\rm g} &= A v_{\rm visc} + 2 B \eta v_{\rm K} \\
  v_{\rm visc} &= -\frac{3}{\Sigma_{\rm g}\sqrt{r}}\fracp{}{r}{\left(\nu \Sigma_{\rm g}\sqrt{r}\right)} \label{eq:vvisc} \\
  \eta v_{\rm K} &= -\frac{c_{\rm s}^2}{2 v_{\rm K}}\frac{\partial \ln P}{\partial \ln r} \\
  v_{\rm drift}^{\rm max} &= \frac{1}{2} B v_{\rm visc} - A \eta v_{\rm K}
\end{align}
with St$_i$ the Stokes number of dust species $i$, $v_{\rm K}$ and $\Omega_{\rm K}$ the Keplerian velocity and frequency, $c_{\rm s}$ the gas sound speed, $P$ the gas pressure, and $\nu = \alpha c_{\rm s}^2/\Omega_{\rm K}$ the gas viscosity for dimensionless $\alpha < 1$. Radial drift is fastest for St$_i$ of order unity. In our \texttt{Fid} models, accretion is driven by a magnetized disk wind and not by shear viscosity, and we therefore set $\alpha = 0$.
\footnote{Technically \texttt{Model C-i} of \citetalias{Kunitomo2020} on which our \texttt{Fid} models are based has a non-zero radial-azimuthal stress $\alpha = 8 \times 10^{-5}$. We nevertheless set $\alpha=0$ in \texttt{Fid} runs to more accurately model MDW disks whose midplanes are strictly laminar in the absence of gas-dust interactions (e.g.~\citealt{Bai2013a}). The inconsistency is small because accretion in \texttt{Model C-i} is driven largely by the MDW torque and not by the prescribed $\alpha$. Later when we discuss equation (\ref{eq:dust_diffusion}) we will account for a non-zero dust diffusivity caused by streaming instability turbulence.} 
For our \texttt{PEW} model, we set, following \citetalias{Kunitomo2020}, $\alpha = 8\times 10^{-5}$ (characterizing gas viscosity of unspecified origin).

The back-reaction coefficients $A$ and $B$, defined in Appendix A of \citet{Grate2019}, quantify how much momentum is transferred from dust to gas via drag. Although including such feedback is formally inconsistent with our use of gas disk models from \citetalias{Kunitomo2020} who neglect such effects, the feedback is never strong enough to significantly alter the gas disk evolution, since the local dust-to-gas ratio is always $\ll 1$ in our models. 

The conversion of dust with surface density $\Sigma_{\rm d,i}$ into planetesimals with surface density $\Sigma_{\rm pl}$ is accounted for in Equation (\ref{eq:sigma_d}) by the source (actually sink) term $S_{\rm ext,i}$:
\begin{equation} \label{eq:sink}
  \sum_i S_{\rm ext,i} = - \sum_i \zeta \Sigma_{\rm d,i} \, \mathrm{St}_i \, \Omega_{\rm K} \equiv -\fracp{\Sigma_{\rm pl}}{t}
  \,.
\end{equation}
We prescribe a formation efficiency $\zeta$ motivated by the streaming instability (SI) simulations of \citet{Li_Chiang_2025} that formed CCKBOs under gas-depleted conditions 
\rlr{(see also discussions of efficiency in e.g. \citealt{Drazkowska2016,Carrera2017,Lau2025})}.  Planetesimals form locally if the following two conditions are met.\footnote{Later in Section \ref{subsec:fid_pi} we will describe a variant of our fiducial run (\texttt{Run Fid-$\Pi$}) that imposes an additional criterion for planetesimal formation to further assess robustness.} 
First, \rlr{motivated by previous works showing that strong clumping by the streaming instability typically requires both aerodynamically large grains and a midplane dust-to-gas density ratio larger than unity \citep[e.g.][]{Youdin_Goodman_2005,LY21}, we require} the midplane density summed over all particles with St$_i > 0.1$ to exceed the midplane gas density:
\begin{align}\label{eq:condition1}
  \frac{\rho_{\rm d,mid}(\mathrm{St} > 0.1)}{\rho_{\rm g,mid}} > 1 
\end{align}
where $\rho_{\rm d,mid,i}=\Sigma_{\rm d,i}/(\sqrt{2\pi} H_{\rm d,i})$, $\rho_{\rm g,mid}= \Sigma_{\rm g}/(\sqrt{2\pi} H_{\rm g})$, $H_{\rm d,i}=\sqrt{\delta/(\delta + {\rm St}_{\rm i})}H_{\rm g}$, and $\delta$ is a dimensionless dust diffusivity. For ${\rm St} \sim 1$, we take $\delta \sim 10^{-4}$ so that $H_{\rm d} \sim 10^{-2}H_{\rm g}$ (see below for more details on $\delta$ and our choices for this parameter). 

Our second planetesimal formation criterion is that there must be sufficient background gas that SI-induced clumping causes the local volumetric density (dominated by solids) to robustly exceed the Roche density. From the simulations of \citet{Li_Chiang_2025}, this second criterion can be translated roughly as
\begin{align}\label{eq:condition2}
  \Sigma_{\rm g}/\Sigma_{\rm MMSN} > 0.03
\end{align}
where $\Sigma_{\rm MMSN} = 2200 \,(r/\text{au})^{-3/2}$ g cm$^{-2}$ \citep{Chiang2010}.  The factor of $0.03$ corresponds to a Toomre $Q$ of $\sim$530 for the gas disk, or equivalently a Roche-to-midplane gas density ratio of $\sim$3000. In other words, we are assuming that streaming turbulence, when fully developed because  criterion (\ref{eq:condition1}) is satisfied, clumps solids so that their density increases by a factor of $\gtrsim$ 3000, enough to trigger gravitational instability when criterion (\ref{eq:condition2}) is satisfied. We experiment with replacing the factor of 0.03 in equation (\ref{eq:condition2}) with 0.003 (\texttt{Run Fid-$\Sigma$}).

If both conditions (\ref{eq:condition1}) and (\ref{eq:condition2}) are satisfied in a given radial bin, then the formation efficiency $\zeta=10^{-4}$ (measured per dynamical time $\Omega^{-1}$ for St = 1 particles); otherwise $\zeta = 0$. Our fiducial $\zeta=10^{-4}$ is motivated by \texttt{Runs C} and \texttt{D} of \citet{Li_Chiang_2025}, in which $\sim$1\% of the dust mass is converted into planetesimals over a timescale of $\sim$$100/\Omega$. Larger values, $\zeta \sim 10^{-3}$--$10^{-2}$, could be inferred from previous studies that intentionally adopted parameters favorable to strong clumping and efficient planetesimal formation \citep[e.g.][]{Simon2016,Li2019}.  We regard $\zeta=10^{-4}$ as more representative of conditions near the onset of planetesimal formation \citep{LY21, Lim2026}.  The value of $\zeta = 10^{-4}$ is nevertheless large enough that most if not all of the ${\rm St}>0.1$ dust particles that satisfy our criteria within a given radial bin do eventually form planetesimals; i.e., the conversion of eligible large dust particles into planetesimals is nearly 100\%. We have verified that increasing $\zeta$ to $10^{-3}$ does not alter our results.

The remaining piece of equation (\ref{eq:sigma_d}) to be explained is the second term in the square brackets, governing turbulent diffusion of dust in gas in the radial direction. The dust diffusion coefficient $D_i$ is given by
\begin{equation} \label{eq:dust_diffusion}
  D_{i} = \frac{\delta c_{\rm s}^2}{\Omega_{\rm K}}\frac{1}{1 + {\rm St}_i^2}
\end{equation}
where $\delta$ is a dimensionless  diffusivity. For simplicity in this work, we ignore (typically order-unity) differences in diffusivities in different directions.  For our \texttt{PEW} model, we set $\delta = 8 \times 10^{-5}$, identical to our turbulent gas diffusivity $\alpha$.  In our \texttt{Fid} models for which $\alpha = 0$, we assume that dust diffusion is dominated by SI turbulence. We utilize measurements of such turbulence made by \citet{LY21}, specifically the geometric mean of the radial and vertical diffusivities as a function of St in their Figure 6 (i.e.~the boundary separating their yellow and pink regimes). When evaluating this mean curve, we use from our simulations the mass-weighted Stokes number
\begin{equation}
  \langle \mathrm{St} \rangle \equiv \frac{\sum_i \Sigma_{\rm d,i} \mathrm{St}_i}{\Sigma_{\rm d}} \,.
\end{equation}
The \citet{LY21} mean curve is followed exactly for $0.01 \leqslant \mwSt \leqslant 0.5$, and simplified at its endpoints, where we assign $\delta = 10^{-6}$ for $\mwSt \leqslant 0.01$ and $\delta = 10^{-4}$ for $\mwSt \geqslant 0.5$.  For reference, $\delta \simeq 4\times10^{-5}$ at $\mwSt \simeq 0.1$.

At every $r$, the distribution of Stokes numbers evolves according to the Smoluchowski equation for the distribution of individual dust particle masses: 
\begin{align}
  \fracp{}{t} N(m) = \int\int M(m, m', m'') N(m') N(m'') \textnormal{d}m' \textnormal{d}m'' \label{eq:smoluchowski}
\end{align}
where $m$ is the mass of an individual dust particle, and $N(m)\equiv \textnormal{d}N/\textnormal{d}m$ is the differential dust surface number density.  The collisional kernel 
\begin{equation}\label{eq:kernel}
  \begin{split}
      M&(m, m', m'') = \\
      &\frac{1}{2} K(m', m'')\cdot \Kdelta(m'+m''-m) - K(m', m'')\cdot \Kdelta(m''-m) \\
      &+ \frac{1}{2} L(m', m'')\cdot S(m, m', m'') - L(m', m'')\cdot \Kdelta(m-m'')
  \end{split}
\end{equation}
where $K$ and $L$ are the coagulation and fragmentation kernels, respectively, $S$ is the distribution of fragments produced by a complete fragmentation, and $\Kdelta$ is the Dirac delta function (which reduces to the Kronecker delta upon discretization). Formulas for $K$, $L$, and $S$ are given by \citet[][and references therein]{LiChenLin2022}.  Relative velocities between particles are determined by a combination of Brownian motion, radial and azimuthal drift, vertical settling, and turbulence \citep{Ormel2007,dustpy}, and the fragmentation velocity threshold is set to $10$ m s$^{-1}$, appropriate for icy grains \citep{Birnstiel2016}. Particle mass $m$ is converted to particle radius assuming spherical dust grains of internal density $1.67 \, {\rm g/cm}^3$, and particle radius converted to Stokes number using equation (10) of \citet{dustpy}.

Eqs.~(\ref{eq:sigma_d}) and (\ref{eq:smoluchowski}) are solved together as a single linear system \citep{dustpy,Birnstiel2010}.

\subsection{Boundary conditions for dust evolution, and resolution in radius and time}
\label{subsec:BCs}

We simulate the evolution of dust and its conversion into planetesimals using \texttt{DustPy-GPU} on a domain that extends from $r = 1$ to $300$ au.  At the outer boundary, a constant-gradient condition is imposed. At the inner boundary, we apply a diode condition that permits only dust outflow (out of the domain). Without the diode condition, we would be forced to model how dust flows into the domain from $r < 1$ au, because in the \citetalias{Kunitomo2020} gas disk models that we use, a gas cavity opens at small $r$ whose positive radial pressure gradient transports dust outward.  Since only $\sim$15\% of the dust in our models is initially contained at $r < 1$ au, and since we are more concerned with planetesimal formation in the outer Solar System, we neglect this innermost inflow.

Our radial grid is not uniform but concentrates resolution where most planetesimals form, including in the CCKB region. The grid is constructed by first laying down 80 cells spaced uniformly logarithmically from $r=1$--300 au, and then refining the grid spacing by a factor of 4 between $r = 5$--65 au (with smooth transitions between the coarse and fine grids). The final grid has 196 radial cells in total. (Later we report on the results of resolution tests that further refine the grid spacing between $r = 35$--55 au.) Timesteps taken by \texttt{DustPy-GPU} are adaptive and variable, and generally much smaller than the interval between gas disk snapshots as given by \citetalias{Kunitomo2020}. For example, for \texttt{Run Fid}, \texttt{DustPy-GPU} takes $\sim$474k integration steps from $t = 0$ to 6 Myr.

\section{Results}
\label{sec:results}

Section \ref{subsec:C-i} reviews the results from our fiducial \texttt{Fid} models which include both MDW and PEW. There we discuss how outcomes depend on input parameters (e.g.~the initial global dust-to-gas ratio $Z$), and in Section \ref{subsec:fid_pi}, planetesimal formation criteria. Section \ref{subsec:A-i} describes the results for our \texttt{PEW} model where gas accretes by viscously diffusing ($\alpha \neq 0$), and escapes by PEW but not MDW.

\subsection{\texttt{Fid} models (MDW + PEW)}
\label{subsec:C-i}

\begin{figure*}
  \centering
  \includegraphics[width=0.8\linewidth]{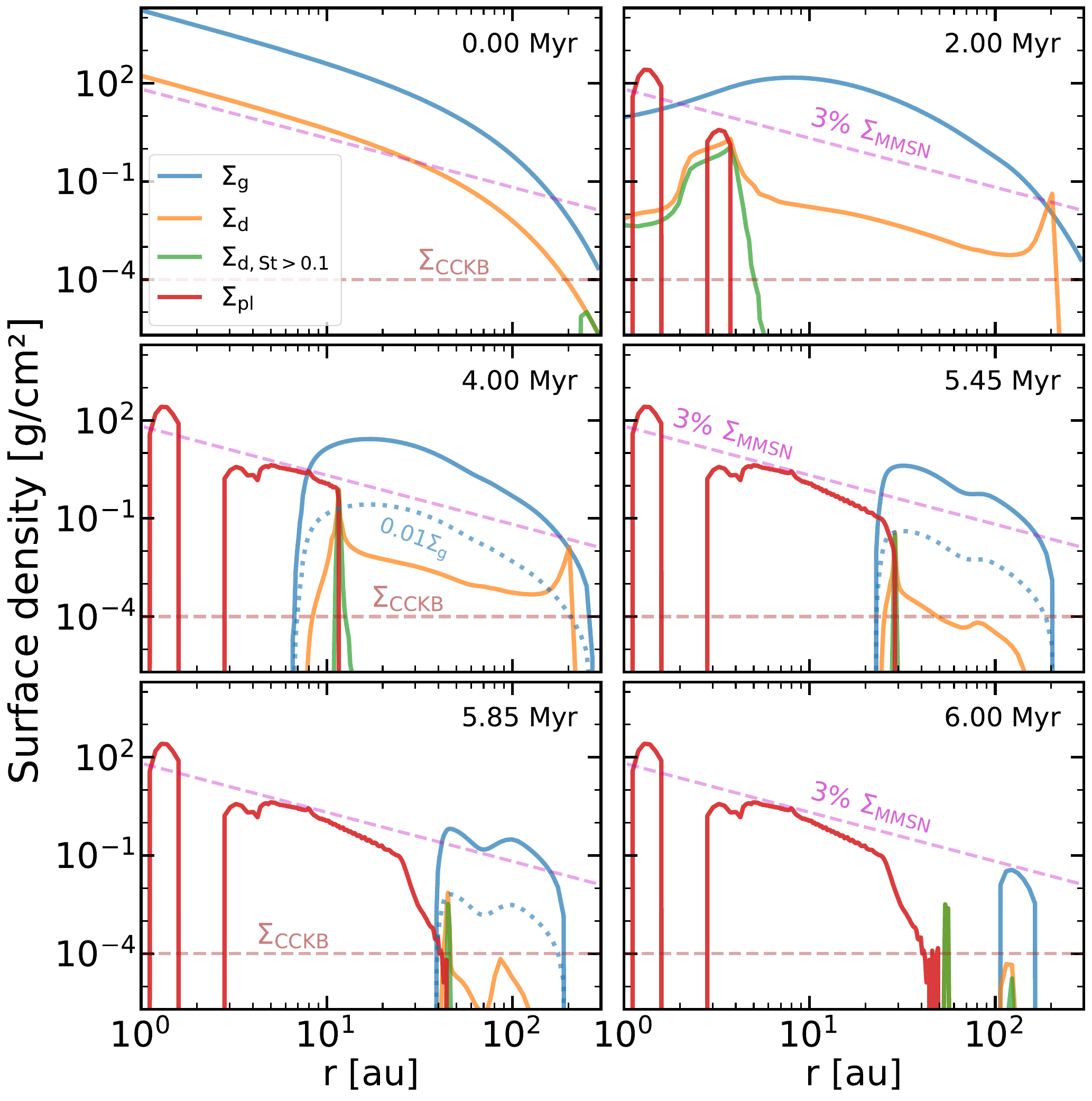}
  \caption{Snapshots of surface density profiles from \texttt{Run Fid}, which includes a magnetically driven disk wind (MDW), a photoevaporative wind (PEW), and particle diffusivities from gas-dust interactions. Gas profiles (blue) are taken from \texttt{Model C-i} of \citetalias{Kunitomo2020}, and  curves for dust (orange), large-sized dust (green), and planetesimals (red) are computed in-house using \texttt{DustPy-GPU}. The ``main dust spike'' (coincident orange and green local maxima first appearing here at $t = 2$ Myr) propagates outward, following the expanding inner edge of the gas disk as it empties from the inside out. The main spike leaves in its wake a ``main disk'' of planetesimals that by $t = 6$ Myr extends from $r \simeq 3$ au to $r \simeq 50$ au. Where the planetesimal disk cuts off at 40--50 au, its surface density $\Sigma_{\rm pl}$ matches that of the Cold Classical Kuiper belt today, $\Sigma_{\rm CCKB}$ (dashed horizontal line). Planetesimal  formation according to our prescription requires enough local gas that clumping of dust from streaming turbulence can drive local dust densities comfortably above the Roche density (eqn.~\ref{eq:condition2}). The required minimum gas surface density is estimated to be $0.03 \, \Sigma_{\rm MMSN}$ \citep{Li_Chiang_2025}, shown as a dashed diagonal line. A further criterion in eqn.~(\ref{eq:condition1}) for planetesimal formation approximately translates into $\Sigma_{\rm d,St>0.1}$ exceeding $0.01 \Sigma_{\rm g}$. The latter is shown in select panels as a dotted blue curve, and is exceeded by the main dust spike by factors of several at early times, but only marginally at late times.  \rlr{The red spike of planetesimals at $r \approx 1$ au lies close enough to our simulation inner boundary and appears early enough in the simulation that it may not be a robust feature (see footnote \ref{foot:spikes}). Animations of this figure and others are available at \href{https://www.rixinli.com/cckb}{rixinli.com/CCKB}.}
  \label{fig:run_Fid}}
\end{figure*}

Figure \ref{fig:run_Fid} shows the evolution of surface densities for gas ($\Sigma_{\rm g}$), dust ($\Sigma_{\rm d}$), large dust ($\Sigma_{\rm d,St > 0.1}$), and planetesimals ($\Sigma_{\rm pl}$) for \texttt{Run Fid}. In a series of snapshots from $t = 0$ to $6$ Myr, we see the development of a ``main dust spike'' where the dust density $\Sigma_{\rm d}$ (orange curve) attains a local maximum, starting at small $r$ and propagating outward with time. The dust spike is composed largely of large particles ($\Sigma_{\rm d} \simeq \Sigma_{\rm d,St>0.1}$; green and orange spikes match). Dust grains start small with low diffusivities, which permit them to grow; thereafter they drift in the radial direction. In the inner disk, dust drifts outward because of the positive radial pressure gradient (blue curve showing the opening of an inner cavity). The inner gas disk drains from the inside out because the MDW depletes gas at smaller $r$ faster (\citetalias{Kunitomo2020} eqn.~29), and because the PEW, which is stronger at larger $r$, starves the inner disk (see also \citealt{Alexander2014}). As dust grains drift outward through the inner disk, and inward through the outer disk where the pressure gradient is negative, they pile up in the main spike, which moves outward with the expanding edge of the gas cavity.
\footnote{In addition to the main dust spike, there are two other ``boundary spikes'' in Fig.~\ref{fig:run_Fid}, one in planetesimals at the inner simulation boundary (red spike at $r \simeq 1$ au), and another near the outer boundary (orange spike at $r \simeq 200$ au). Both appear early at $t < 1.5$ Myr, the inner spike because grains grow quickly in the dense innermost regions and pile up as they drift radially outward, and the outer spike because the low gas density in the outermost disk leads to order-unity Stokes numbers for the particles, which  pile up from fast inward drift before they can grow. The outer spike eventually dissolves in \texttt{Run Fid}. Because these spikes are located close to our simulation boundaries and appear early, they are sensitive to our assumed initial and boundary conditions.\label{foot:spikes}} 
The enhancement of density in the main spike helps dust there to grow to St $> 0.1$.

The main dust spike is where planetesimals form. A wave of planetesimal formation moves outward with the spike, creating a permanent disk of solids in its wake (red curve in Fig.~\ref{fig:run_Fid}). Initially, at early $t$ and small $r$, planetesimals form out of dust with order-unity efficiency; the leading edge of the planetesimal disk has a surface density $\Sigma_{\rm pl}$ comparable in magnitude to the dust spike surface density $\Sigma_{\rm d} \simeq \Sigma_{\rm d,St>0.1}$ (see the convergence of red, green, and orange curves in, e.g., the $t=4$ Myr snapshot at $r \simeq 10$ au). Later, at $t \gtrsim 5.4$ Myr and $r \gtrsim 25$ au, $\Sigma_{\rm pl}$ at the leading edge falls below $\Sigma_{\rm d,St>0.1}$, and the planetesimal surface density drops off more rapidly with radius. The steeper decline in $\Sigma_{\rm pl}$ occurs when so much of the disk's total mass in dust has been consumed by prior planetesimal formation that the criterion in eqn.~(\ref{eq:condition1}) becomes only marginally satisfied inside the spike. In Fig.~\ref{fig:run_Fid} we show in the $t = 5.45$ and 5.85 Myr panels  how $\Sigma_{\rm d,St>0.1}$ only just intersects $0.01 \times \Sigma_{\rm g}$ (dotted blue curves). As the dust scale height $H_{\rm d}$ is about 0.01 $\times$ the gas scale height $H_{\rm g}$ (see text after eqn.~\ref{eq:condition1}), in both snapshots the midplane dust density $\rho_{\rm d,mid}({\rm St} > 0.1)$ is close to the midplane gas density $\rho_{\rm g,mid}$, and eqn.~(\ref{eq:condition1}) is satisfied by only a small margin. The small excess density $\rho_{\rm d,mid}({\rm St} > 0.1) - \rho_{\rm g,mid}$ is converted into planetesimals with small $\Sigma_{\rm pl}$, and $\Sigma_{\rm d,St>0.1}$ becomes capped at $\sim$$0.01 \,\Sigma_{\rm g}$. Contrast these throttled states with the earlier snapshot at $t = 4$ Myr when dust is still sufficiently plentiful and collects sufficiently rapidly that $\Sigma_{\rm d,St>0.1}> 0.01\,\Sigma_{\rm g}$ and planetesimals form in abundance.

\begin{figure*}
  \centering
  \includegraphics[width=0.9\linewidth]{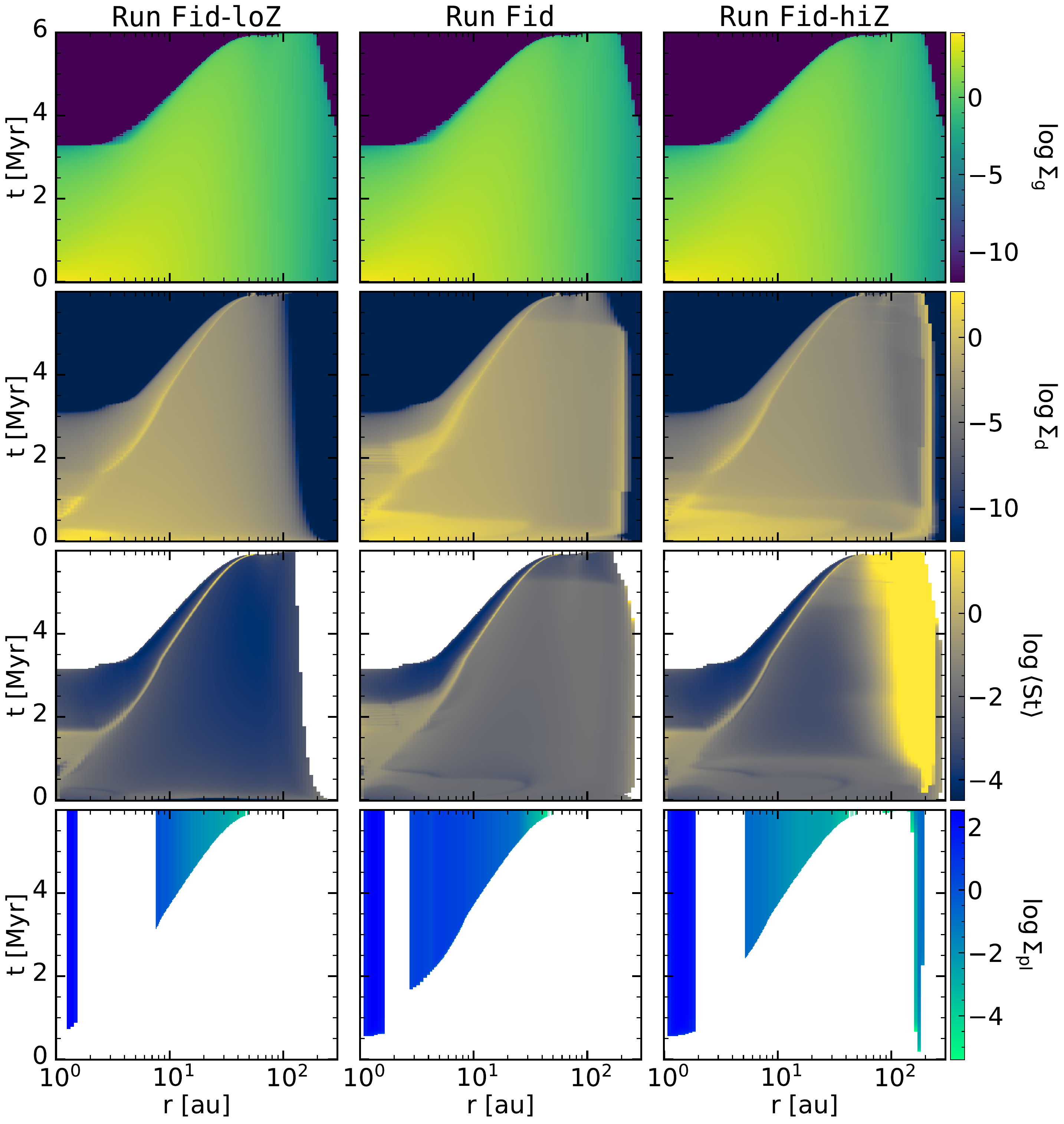}
  \caption{Space-time diagrams showing the evolution of surface densities (in units of g/cm$^2$) and the mass-weighted Stokes number $\langle \mathrm{St} \rangle$ of dust particles in \texttt{Runs Fid-loZ} (left column), \texttt{Fid} (middle column), and \texttt{Fid-hiZ} (right column).  The bright yellow streamer in the middle two rows traces the ``main spike'' of dust particles that moves outward following the expanding inner edge of the gas disk. Planetesimal formation inside the main spike creates a ``main planetesimal disk'' (triangles in the bottom row) that truncates at its outer edge between 40--50 au, coincident with the location of the CCKB, and with about the right surface density of $\Sigma_{\rm CCKB} = 10^{-4}$ g/cm$^2$.
  \label{fig:run_Fid-st}}
\end{figure*}

As the planetesimal disk thins out, $\Sigma_{\rm pl}$ eventually crosses the present-day CCKB surface density  $\Sigma_{\rm CCKB} = 10^{-4}$ g/cm$^2$.  In \texttt{Run Fid}, the crossing occurs at $r \simeq 42$ au, fortuitously close to where the CCKB resides. From $r \simeq 42$ au to 50 au, planetesimal formation is ``touch-and-go'', as the criteria in eqs.~(\ref{eq:condition1}) and (\ref{eq:condition2}) are only intermittently satisfied against a diminishing dust mass and a gas disk eroding increasingly quickly; see the space-time diagram for $\Sigma_{\rm g}$ in Figure \ref{fig:run_Fid-st}, showing how the gas cavity expands especially fast in the last $\sim$0.5 Myr.  So far as we can tell, the intermittency is not due to lack of spatial resolution; further refining the grid spacing between $r=35$--55 au by a factor of 4 (so that grid cells are $\sim$0.2 au wide at $r=45$ au) does not change our results qualitatively. 

\newcommand{\savedDblFloatPageFraction}{}
\let\savedDblFloatPageFraction\dblfloatpagefraction
\renewcommand{\dblfloatpagefraction}{0.95}
\begin{figure*}
  \centering
  \includegraphics[width=0.8\linewidth]{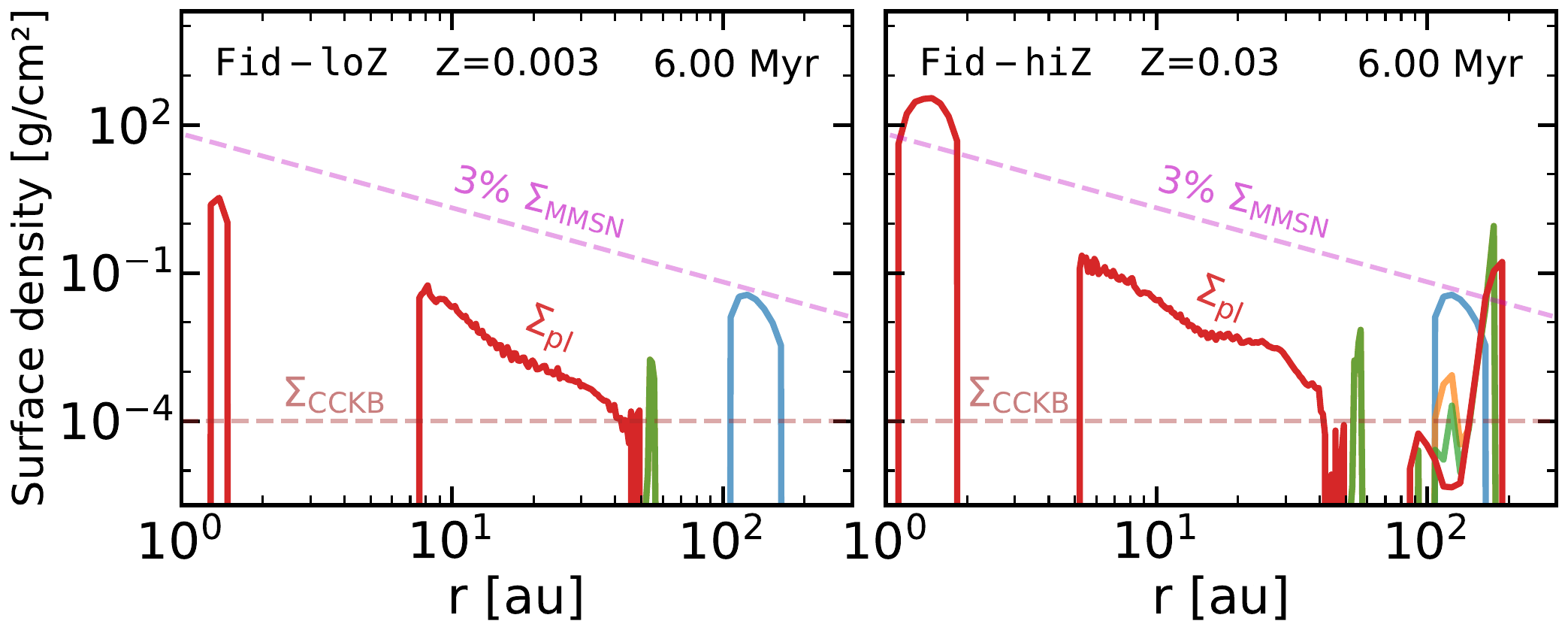}
  \caption{Same as Fig.~\ref{fig:run_Fid} but showing the final $t = 6$ Myr surface density snapshots for \texttt{Runs Fid-loZ} (left; initial global dust-to-gas ratio $Z = 0.003$) and \texttt{Fid-hiZ} (right; $Z = 0.03$). As $Z$ varies, the main disk of planetesimals (red curve in the middle of each panel) varies in the radius of its inner edge and in total mass, while the outer radius (40--50 au) and outer surface density ($\Sigma_{\rm pl} \sim 10^{-4}$ g/cm$^2$) remain similar to those of the CCKB.
  \label{fig:run_Fid-varZ-final}}
\end{figure*}
\afterpage{\renewcommand{\dblfloatpagefraction}{\savedDblFloatPageFraction}}

Experimentation shows that the formation of a ``main disk'' of planetesimals extending from $r \simeq 3$--8 au to $r \simeq 40$--50 au is robust against variations in:
\begin{enumerate}[label=\roman*)]
  \item initial dust grain sizes (varied from 0.5 to 100 $\mu$m)
  \item the planetesimal formation efficiency $\zeta$ (changed from our default $10^{-4}$ to $10^{-3}$; eqn.~\ref{eq:sink})
  \item the minimum $\Sigma_{\rm g}$ needed for SI-induced clumps to exceed the Roche density and form planetesimals (changed from our default $0.03 \,\Sigma_{\rm MMSN}$ to $0.003 \,\Sigma_{\rm MMSN}$; eqn.~\ref{eq:condition2}; see \texttt{Run Fid-$\Sigma$} in Table \ref{tab:paras})
  \item the initial global dust-to-gas ratio $Z$ (varied from 0.003 to 0.03).
\end{enumerate}
Regarding item (iv), Figure \ref{fig:run_Fid-varZ-final} shows final snapshots of surface densities in \texttt{Runs Fid-loZ} and \texttt{Fid-hiZ}. As compared to \texttt{Fid}, the main disk mass is a factor of 50--100 lower in \texttt{Fid-loZ} and \texttt{Fid-hiZ} (Table \ref{tab:paras}, column 9). The former starts with fewer solids which results in less growth, and the latter exhibits early and efficient planetesimal formation at the inner simulation boundary (see footnote \ref{foot:spikes}) which reduces the amount of dust that drifts outward to create the main disk.  Both the innermost planetesimal ring at $r \simeq 1$ au and the outermost ring at $r \simeq 100$--200 au form early in \texttt{Run Fid-hiZ} (Fig.~\ref{fig:run_Fid-st}), and both may not be robust outcomes insofar as they depend on our assumed initial and boundary conditions. While the total mass in planetesimals in the main disk is sensitive to how much mass is sequestered by the innermost ring, the outer radius of the main disk appears relatively unaffected; $\Sigma_{\rm pl}$ crosses $\Sigma_{\rm CCKB}$ at $r\simeq 40$--50 au regardless of what is happening at the simulation boundaries at early times.

One concern with our results is that planetesimals form in the main dust spike where the gas pressure gradient is smallest and where streaming motions may be too weak to trigger the SI in the way we have prescribed (\citealt{Li_Chiang_2025}; \citealt{Lee2022}). We address this issue in the following Section \ref{subsec:fid_pi}.

\subsubsection{\texttt{Fid-$\Pi$}}
\label{subsec:fid_pi}

As discussed above, the \texttt{Fid} models form planetesimals/CKBOs ostensibly via the SI inside a (moving) gas pressure bump. A potential problem with this scenario is that it might conflict with the findings of \citeauthor{Li_Chiang_2025} (\citeyear{Li_Chiang_2025}; see also \citealt{Lee2022}), who document how planetesimal formation inside a (static) pressure bump can proceed differently from planetesimal formation induced by the SI.  Inside a bump, the pressure gradient is small and by extension gas-dust relative streaming velocities are small, suppressing the SI. The bump simulations of \citet{Li_Chiang_2025} fail to form planetesimals when gas turbulence is imposed at levels comparable to those in our present simulations.

To explore the possibility that streaming motions may be too weak inside our gas pressure bump to trigger SI clumping, we introduce an extra third criterion for planetesimal formation. In addition to requiring the conditions in eqs.~(\ref{eq:condition1}) and (\ref{eq:condition2}) to trigger $\zeta > 0$ in a given grid cell, we also require the dimensionless pressure gradient
\begin{align}\label{eq:pi}
  \Pi \equiv \left| \frac{\eta v_{\rm K}}{c_{\rm s}} \right| \equiv \left| \frac{1}{2}\frac{c_{\rm s}}{v_{\rm K}}\fracp{\,\ln{P}}{\,\ln{r}} \right| > \Pi^\ast
\end{align}
for some threshold $\Pi^\ast$.
\footnote{Where $\Pi < \Pi^\ast$, we still allow for planetesimal formation if the total midplane dust density $\rho_{\rm d,mid} > 4 \times$ the Roche density, with $\zeta = 10^{-2}$. This high-efficiency channel models formation by direct gravitational instability, as seen in the pressure bump simulations of \citeauthor{Li_Chiang_2025} (\citeyear{Li_Chiang_2025}, their \texttt{Runs Bu} and Fig.~9). In practice, however, this possibility of direct collapse is never realized in our calculations.} 
The sign of the pressure gradient is ignored since the SI operates whether dust rotates faster or slower than gas; what matters is that the relative streaming motion be non-zero. For reference, the smooth disk (no bump) SI simulations that reproduced the CCKB in \citet{Li_Chiang_2025} have $\Pi = 0.056$. \citet{Bai2010b} found SI clumping down to $\Pi = 0.025$.  With single-species ${\rm St} = 0.3$ experiments in 2D shearing boxes, we found SI clumping down to $\Pi = 0.01$ (data not shown).

\begin{figure*}
  \centering
  \includegraphics[width=0.8\linewidth]{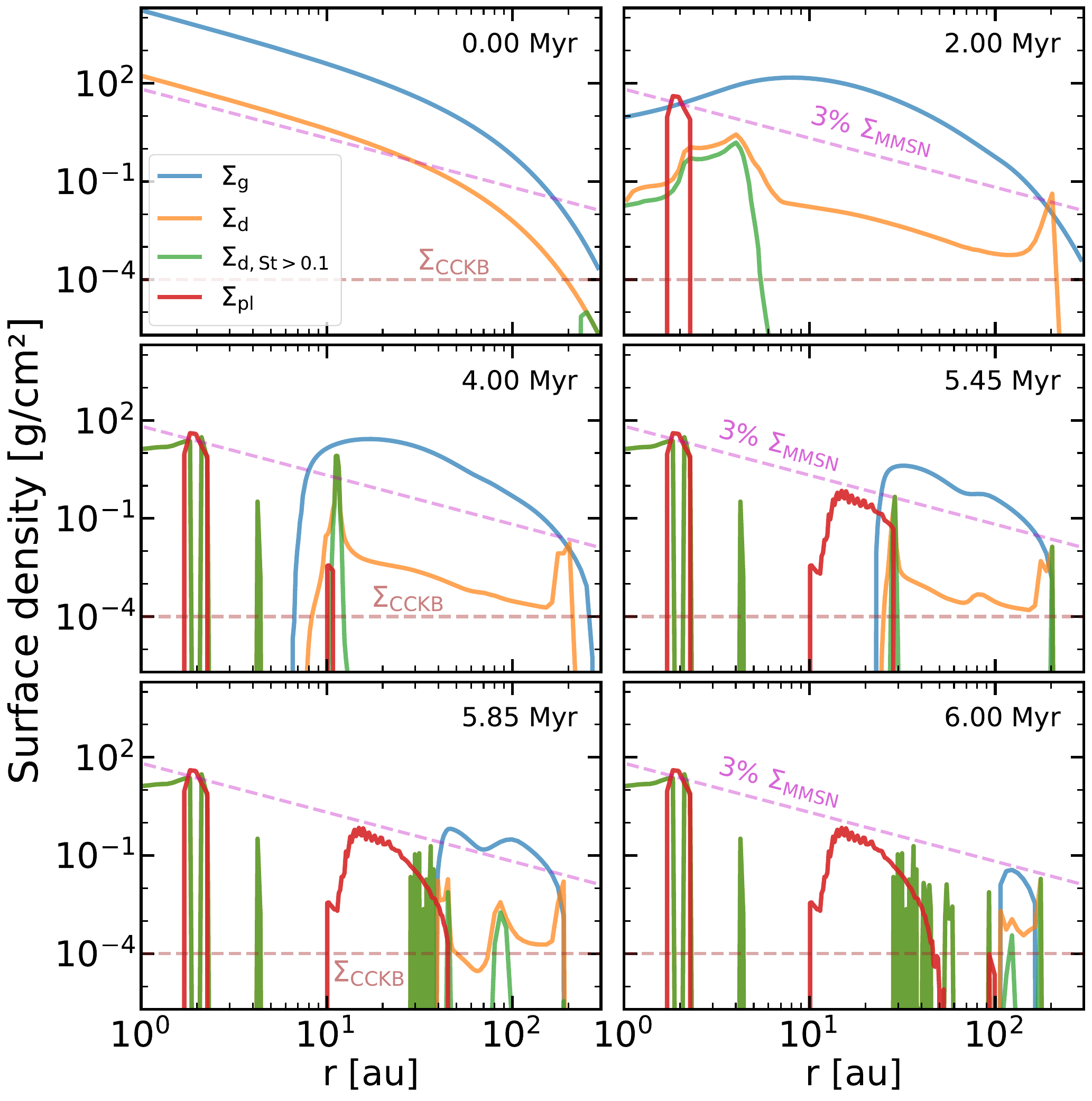}
  \caption{Same surface density snapshot sequence as in Fig.~\ref{fig:run_Fid} but for \texttt{Run Fid-$\Pi$} which imposes an additional constraint on planetesimal formation, that the dimensionless pressure gradient $\Pi > \Pi^\ast = 0.01$ (eqn.~\ref{eq:pi}). This constraint prevents planetesimals from forming at the exact center of the gas pressure bump (= exact center of the main dust spike), where relative streaming motions between dust and gas may be too weak for SI turbulence to fully develop \citep{Li_Chiang_2025}.  Accordingly, fewer planetesimals form in \texttt{Fid-$\Pi$} as compared to \texttt{Fid}. Despite the added constraint, the planetesimal disk truncates at its outer edge approximately the same way between \texttt{Fid-$\Pi$} and other \texttt{Fid} runs, with properties similar to those of the CCKB ($\Sigma_{\rm pl} \sim \Sigma_{\rm CCKB}$ at $r\simeq 40$--50 au).
  \label{fig:fid_pi}}
\end{figure*}

We re-run our \texttt{Fid} model with the added criterion in eqn.~(\ref{eq:pi}), experimenting with $\Pi^\ast$ between 0.001 and 0.01.  Results for our most stringent test with $\Pi^\ast = 0.01$ (\texttt{Run Fid-$\Pi$}) are shown in Table \ref{tab:paras} and Figure \ref{fig:fid_pi}.  Increasing $\Pi^\ast$ turns off planetesimal formation in an increasing number of grid cells centered on the main spike of large dust particles. As $\Pi^\ast$ increases from 0 to 0.01, the total mass in planetesimals that are formed ($M_{\rm pl,total}$ in Table \ref{tab:paras}) decreases by a factor of $\sim$3, and the mass in leftover dust ($M_{\rm dust,final}$) increases correspondingly. Nevertheless, the truncation of the main planetesimal disk at $r \simeq 40$--50 au is remarkably insensitive to $\Pi^\ast$. 

Our $\Pi^\ast$ experiments show that planetesimals are created not at the pressure bump center but rather in the bump wings, where dust is still drifting relative to gas. Thus we argue our CCKB formation scenario is consistent with the ``drift-and-form'' scenario of \citet{Li_Chiang_2025}.

\subsection{\texttt{PEW}-only model}
\label{subsec:A-i}

\begin{figure*}
  \centering
  \includegraphics[width=0.75\linewidth]{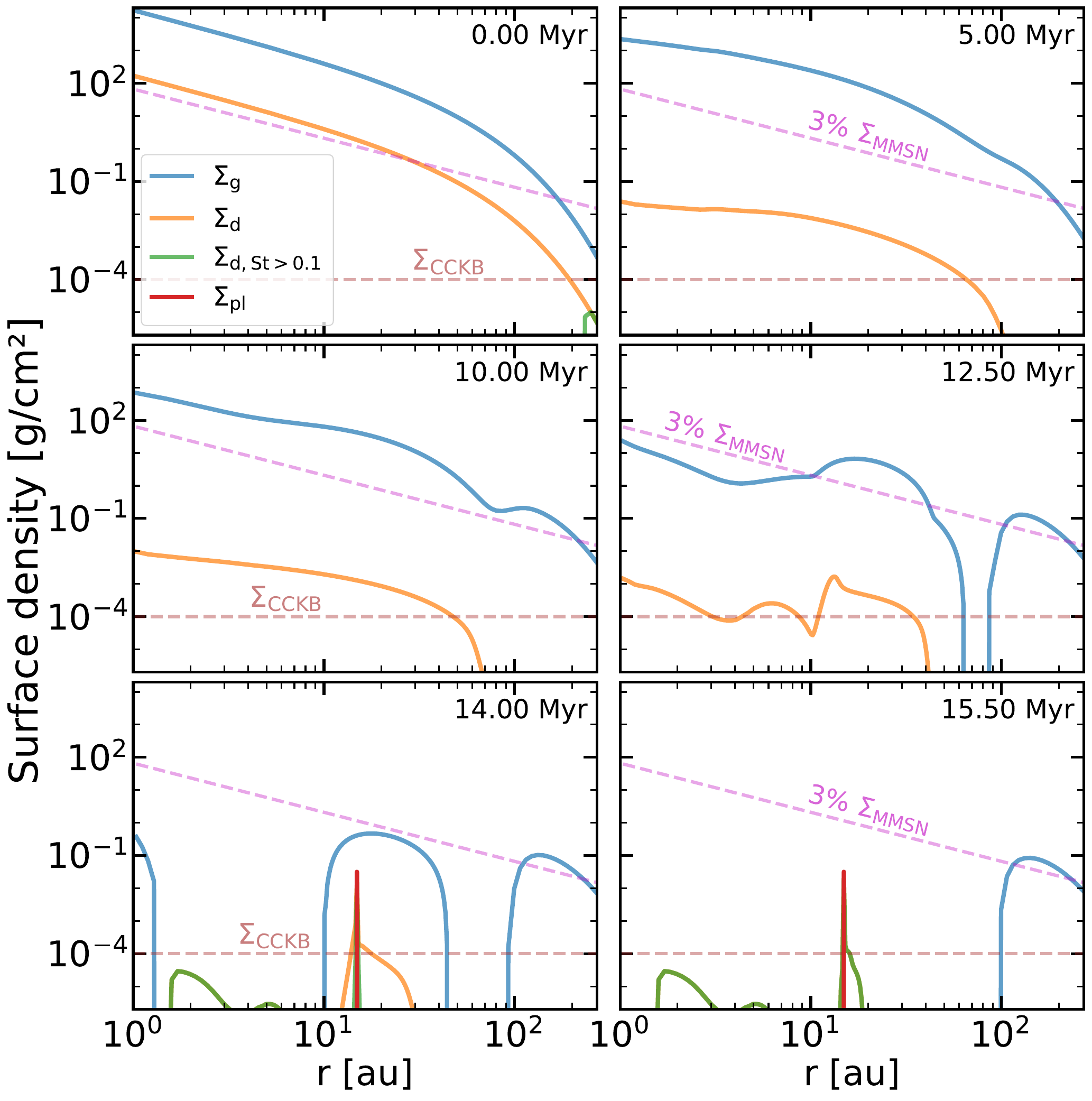}
  \caption{Surface density snapshot sequence for \texttt{Run PEW} which accounts for a photoevaporative wind and replaces the \texttt{Fid} MDW accretion torque with an $\alpha = 8 \times 10^{-5}$ shear viscosity. Unlike the \texttt{Fid} models where the inner gas disk immediately starts clearing from the inside out to create a pressure bump at $t \lesssim 1$ Myr, here for \texttt{PEW} a pressure bump does not appear until after $\sim$10 Myr, when the viscous accretion rate in the inner disk finally drops below the PEW mass loss rate and the inner disk starves   (e.g.~\citealt{Alexander2014}).     At earlier times, the gas disk surface density declines monotonically outward, and most dust particles are lost to inward radial drift, failing to form planetesimals that might grow to become planets or Kuiper belt objects. The red spike of planetesimals appearing at $r \simeq 15$ au contains $\sim$0.03 $M_\oplus$; even this population of planetesimals is not a robust outcome because it forms inside a single radial grid cell, and is eliminated when a $\Pi^\ast = 0.01$ condition for planetesimal formation is imposed (eqn.~\ref{eq:pi}). See also Figure \ref{fig:run_PEW-st} which includes data on particle Stokes numbers. 
  \label{fig:run_PEW}}
\end{figure*}

\begin{figure*}
  \centering
  \includegraphics[width=\linewidth]{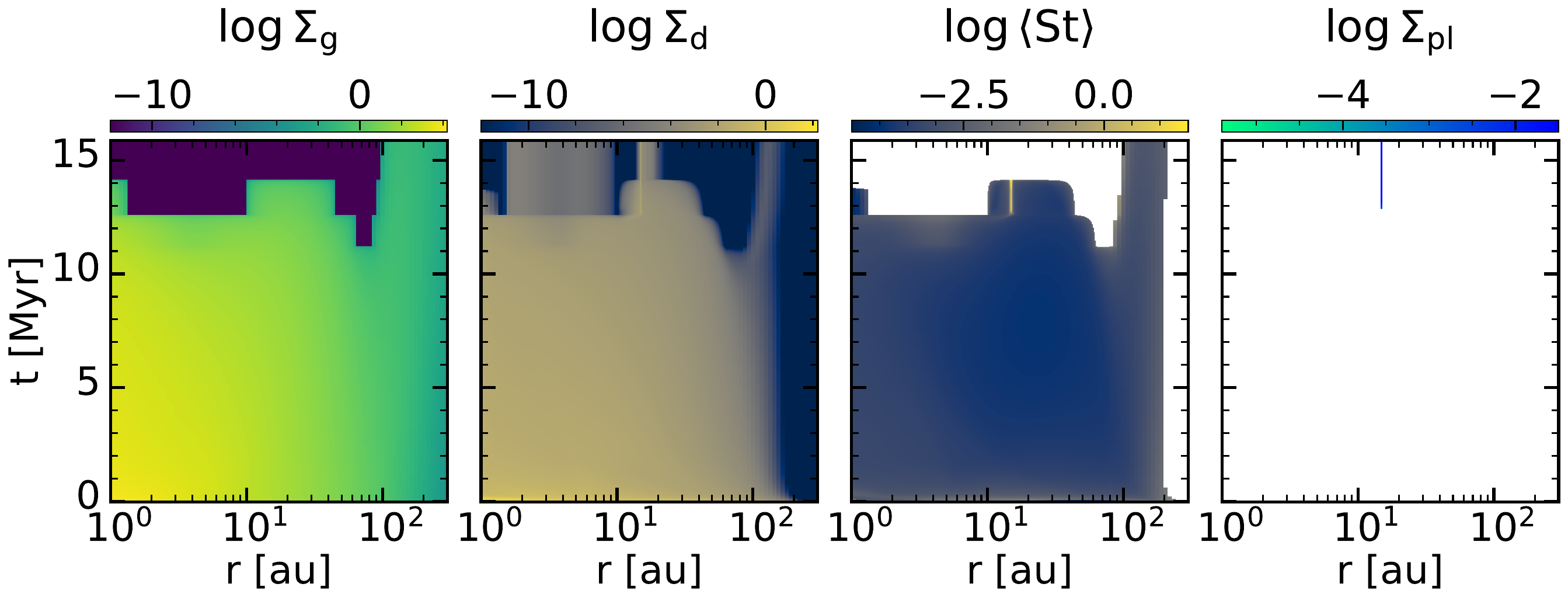}
  \caption{Space-time diagrams for \texttt{Run PEW} showing the evolution of surface densities (in units of g/cm$^2$) and the mass-weighted Stokes number $\langle \mathrm{St} \rangle$ of dust particles. At $t \lesssim 10$ Myr, prior to the clearing of the inner gas disk, dust particles hardly grow; $\langle \mathrm{St} \rangle < 10^{-2}$. At $\sim$12 Myr, a gas pressure bump and attendant dust spike appear at $r \simeq 15$ au, enabling planetesimals to form, but only in a single radial grid cell.
  \label{fig:run_PEW-st}}
\end{figure*}

Figures \ref{fig:run_PEW} and \ref{fig:run_PEW-st} show results for \texttt{Run PEW} where the disk disperses from a combination of an $\alpha$ shear viscosity and  photoevaporation. While the disk drains from the inside out in both \texttt{Runs PEW} and \texttt{Fid}, it does so more slowly in the former from viscous accretion than in the latter from a magnetized wind. In \texttt{PEW}, an inner gap/cavity in the gas disk does not appear until $t > 10$ Myr, when the viscous accretion rate drops below the photoevaporative mass loss rate and the inner disk becomes starved (e.g.~\citealt{Alexander2014}). Before then, the gas disk has a consistently negative radial pressure gradient, and much of the dust, which does not grow to St $\gtrsim 0.01$ (Fig.~\ref{fig:run_PEW-st}), drifts to $r < 1$ au, out of the simulation domain. What little dust remains in the domain at $t > 10$ Myr is collected by a late-time gas pressure bump into planetesimals of mass $\sim$0.03 $M_\oplus$ at $r \simeq 15$ au. The formation of this planetesimal population is not robust because it occurs in only a single radial grid cell; indeed, imposing an extra $\Pi^\ast = 0.01$ condition for planetesimal formation (eqn.~\ref{eq:pi}) prevents these and any other planetesimals from forming. \texttt{Run PEW} fails to reproduce any aspect of the Solar System, including the CCKB.

\section{Summary and Discussion}
\label{sec:summary}

We have developed an origin story for Cold Classical Kuiper belt objects (CCKBOs), which are considered first-generation planetesimals untouched by the upheavals experienced by the rest of the Solar System. In our scenario, the CCKB forms as the ``last gasp'' of a wave of planetesimal formation that sweeps radially outward from the Sun as the solar nebula empties from the inside out. The wave is carried by the gas disk's expanding inner edge, which presents a pressure maximum collecting dust grains from either side. As dust particles drift toward the maximum, they concentrate and trigger the streaming instability (SI), which clumps dust into planetesimals along the lines of the ``drift-and-form'' scenario proposed in our previous work (\citealt{Li_Chiang_2025}). Massive enough to aerodynamically decouple, the planetesimals so formed are dropped off at the gas disk inner edge; the receding gas paves a disk of planetesimals that fills the cavity of this transitional disk. When nearly all the dust has been used up, the planetesimal disk truncates at its outer edge, forming the ``Kuiper Cliff''.

The details of how the planetesimals and in particular CCKBOs are laid down depend on how the background gas disk disperses.  Photoevaporation driven by X-ray and ultraviolet radiation erodes the outer (more radially distant) regions of an accretion disk faster than the inner regions, preventing the inner disk from being resupplied by the outer; circumstellar gas then clears by accreting from the inside out  (e.g.~\citealt{Alexander2014}). While photoevaporation is necessary for inside-out clearing, the details of the clearing and by extension the speed and amplitude of the wave of planetesimal formation that propagates outward depend on the mechanism of gas accretion (angular momentum transport). In a model where gas accretion proceeds viscously (derived from \texttt{Model A-i} of \citealt{Kunitomo2020} = \citetalias{Kunitomo2020} with an $\alpha = 8 \times 10^{-5}$ shear viscosity), an inner transitional disk cavity takes $\gtrsim 10$ Myr to materialize, by which time practically all the disk's dust has drifted inward unchecked and is unavailable for planetesimal formation outside a few au. \rlr{A similar problem of wholesale dust loss on timescales $< 1$ Myr afflicts the $\alpha$ models of \citet{Marschall2023}.} 

By contrast, in a model where the accretion torque is due to a magnetized wind launched from the disk surface (we have used \texttt{Model C-i} of \citetalias{Kunitomo2020}; see also \citealt{Bai2013a}), the disk evacuates from the inside more rapidly. Abetted by photoevaporation, a magnetic wind drains gas from the inside faster than it does the outside (\citetalias{Kunitomo2020} eqn.~29), opening a cavity whose edge attracts and traps dust, and forms planetesimals on $\gtrsim$ au scales in a Myr or so. There are {\it a priori} reasons to believe that disks accrete by magnetized winds and not by turbulence acting as a shear viscosity (e.g.~\citealt{Bai2013a}; \citealt{Bai2016}), and our CCKB formation calculations support this paradigm. \texttt{Model C-i} is also favored by \citetalias{Kunitomo2020} as best matching disk observations \rlr{(e.g.~\citealt{Mamajek2009,Flaherty2017,Ercolano2017,Pascucci2025})}.

Though \texttt{Model C-i} contains no  information about the Kuiper belt, its use in our \texttt{DustPy-GPU} simulations reproduces remarkably well the present-day properties of the CCKB, namely how it truncates with a surface density of $\sim$$10^{-4}$ g/cm$^2$ at a radius of $\sim$40--50 au. The near-perfect match in radius is to an extent fortuitous, but we would argue that the general outline/framework of our formation scenario is robust. We varied a number of input parameters by a factor of 10 or more (see the list in Section \ref{subsec:C-i}), and experimented also with refining the criteria for planetesimal formation (Section \ref{subsec:fid_pi}), and in all such tests reproduced the CCKB comparably well.  The existence of a sharp radial cut-off in the planetesimal disk is inevitable given a finite amount of raw dust, and the requirement that this dust (and the ambient gas) meet threshold conditions (set by the SI) before planetesimals can form. That the cut-off occurs somewhere in the vicinity of 100 au follows from photoevaporation which removes mass on this radial scale to starve the disk at smaller radii, enabling inside-out clearing by magnetized wind accretion.

From here there are any number of open problems to be pursued:
\begin{enumerate}
\item What factors determine the CCKB's characteristic surface density $\Sigma_{\rm CCKB} \sim 10^{-4}$ g/cm$^2$, \rlr{and what other kinds of planetesimal belts can be generated}? We used a generic model for a dissipating gas disk and happened to reproduce the CCKB surface density over the right range of radii $r\simeq 40$--50 au. \rlr{One should consider variations on this model (e.g.~different field strengths for the magnetized disk wind; different stellar host spectral types with different photoionizing fluxes; e.g.~\citealt{Ooyama2025}) to better understand how exactly these numbers arise, and see also if the diversity of extrasolar debris disk radial geometries and surface densities (e.g.~\citealt{Han2026}) can be reproduced. H.~Johnston and S.~Marino (2026, personal communication) report ongoing work along these lines. While there is a sharply defined outer edge in our planetesimal disk models where the surface density drops to zero, the inner edge of any debris belt that forms out of our planetesimal disk should be determined by planet formation. For example, the inner edge could be set by the chaotic zone of an interior planet (\citealt{Quillen2006}). Or the inner edge could be carved by secular resonances with interior planets, as in the CCKB inner edge at 42 au (\citealt{Duncan1995}). Given how our modeled planetesimal surface density drops steeply over many orders of magnitude at its outer edge (e.g.~Fig.~\ref{fig:run_Fid}), small changes in where we draw the inner edge can result in large variations in the resulting debris disk mass. The steep drop in planetesimals can also explain how Neptune stopped migrating (\citealt{Nesvorny2018}).}

A related issue worth exploring is how the planetesimal surface density $\Sigma_{\rm pl}$ fluctuates with $r$ before dropping to zero (see the jaggedness of the red $\Sigma_{\rm pl}$ profiles in Figures \ref{fig:run_Fid}, \ref{fig:run_Fid-varZ-final}, and \ref{fig:fid_pi}). Presumably the speed with which the gas disk erodes from the inside out plays a role; we would guess that the faster the erosion, the less time dust particles have to concentrate radially before the gas recedes, and the fewer planetesimals form. Thus the intermittency may reflect a neck-and-neck race between wind erosion and the streaming instability.  A challenge for observers would be to map the CCKB with finer resolution along the axes of orbital distance and object size/mass.

\item We should assess more carefully how planetesimals may or may not form within local pressure maxima, i.e.~pressure bumps. Our scenario features a moving pressure bump (the expanding inner edge of the gas disk), and we have posited that planetesimals can form within this bump as long as threshold criteria for the SI are satisfied (eqs.~\ref{eq:condition1} and \ref{eq:condition2}). The problem is that these criteria for the SI derive from simulations of smooth, bump-free disks where radial pressure gradients, as measured by the dimensionless parameter $\Pi$ (eqn.~\ref{eq:pi}), can be substantially larger than inside bumps. Pressure gradients drive the gas-dust relative motions that are at the heart of the SI, and indeed at the bump center where the pressure gradient is zero, planetesimal formation has been found to proceed very differently than via the SI (\citealt{Li_Chiang_2025}). We tried to account for this complication by adding a $\Pi > \Pi^\ast$ criterion for planetesimal formation in select runs, and found no qualitative change to our conclusions, as planetesimals form in the broad wings of the bump and not necessarily at the exact bump center. Nevertheless, more explorations of what happens within bumps would be welcome, including how the motion of the bump as a whole may affect its internal dynamics. The effects of smaller $\Pi$ may be complicated to sort out; e.g.~\citet{Bai2010b} found in 2D axisymmetric simulations of the SI that smaller $\Pi$ actually promoted stronger clumping of dust particles (their Figure 2).

\item The present-day heliocentric orbital eccentricities and inclinations of CCKBOs, though smaller than those of other Kuiper belt populations, are larger than predicted by our calculations which are staged at the time of planetesimal formation. From the SI simulations of \citet{Li_Chiang_2025}, ``primordial''  eccentricities of newly formed CCKBOs could be of order the turbulent streaming velocities normalized by the Keplerian velocity, $e \sim \Pi \times h \sim 0.004$, where $\Pi$ is the dimensionless radial pressure gradient and $h$ is the local gas disk aspect ratio. Primordial inclinations can also be estimated from these simulations as the thickness of the dust particle disk divided by the local orbital radius, $i \sim (H_{\rm p}/H) \times h \sim 0.0005$, where $H_{\rm p}$ and $H$ are the respective particle and gas disk scale heights. These formation-era eccentricities and inclinations are considerably lower than current CCKB eccentricities and inclinations, whose rms values are $e\simeq 0.075$ and $i \simeq 0.045$ (these are the free components not forced by Solar System planets today; \citealt{Gladman_Volk2021}; \citealt{Huang2022}). \rlr{While such eccentricities can be generated by Neptune's 2:1 eccentricity resonance in suitably grainy planet migration histories, analogous inclination resonances are weaker and have not been shown to be effective (\citealt{Nesvorny2015,Nesvorny2018}). Long-distance forcing of $i$ and $e$ by a secular or mean-motion resonance may be necessary, as close encounters with perturbers --- a.k.a.~viscous stirring, both natively by CCKBOs and non-natively by a ``rogue planet'' (\citealt{Huang2022b}) or Neptune itself (\citealt{Dawson2012}) --- would disrupt CCKBO binaries. To generate $i,e \sim 0.1$ from a close encounter would require a velocity impulse of $\sim$100 m/s, larger than the $\lesssim 10$ m/s velocities with which CCKBO binary components orbit each other.}  Reproducing the degree to which the CCKB has been stirred (apparently up to equipartition, insofar as the rms eccentricity today is twice the rms inclination; \citealt{Ida1992}) without disrupting this population wholesale remains a challenge.

\item Widening our scope still further, we should ensure that whatever gas and dust disk evolution scenario works for the CCKB also works for the rest of the Solar System. The mass of $46 M_\oplus$ in the ``main planetesimal disk'' in our fiducial \texttt{Fid} model  (Table \ref{tab:paras}) probably falls short (but only by a factor of $\lesssim 2$) of the mass in heavy (non-H, non-He) elements contained in the gas and ice giants. Part of this shortcoming in the outer Solar System mass budget arises because too many solids collect near or drift out of the inner simulation boundary at 1 au, where our model seems too crude and sensitive to initial and boundary conditions. Another concern is planetary gas accretion; draining the gas disk from the inside out, and forming planetesimals only near the inner edge of the eroding gas disk, limits the time available for planetesimals to assemble into the cores of Jupiter and Saturn, and the time for these cores to accrete gas. If we take \texttt{Run Fid} at face value, Jupiter at 5 au and Saturn at 10 au each have $\sim$1 Myr to form starting from dust particles (Fig.~\ref{fig:run_Fid}). Though not obviously infeasible, such a timeline seems a bit fast \rlr{(cf.~\citealt{Kruijer2017})}.
\end{enumerate}

Of course no single 1D model, including ours, can be expected to capture all transport and growth processes occurring over multiple decades in orbital radius. Our focus here has been on reproducing the CCKB, in the belief that it can test  theories of planetesimal formation. In that regard we hope the basic elements of our CCKB creation story are robust and will survive model revisions. These elements include a gas disk that dissipates from the inside out --- not necessarily over the entire disk, but at least over trans-Neptunian space $\gtrsim$ 30 au; and the streaming instability, triggered in dust particles that drift toward the pressure maximum at the gas disk's inner edge, which itself recedes from the Sun.

\vspace{1in}
\section*{Acknowledgements}

We are indebted to Masanobu Kunitomo for generously providing the data underlying the \citetalias{Kunitomo2020} gas disk models \texttt{C-i} and \texttt{A-i}. \rlr{We also thank Anders Johansen, Heather Johnston, Tommy Lau, Sebasti\'an Marino, David Nesvorn\'y, and Andrew Youdin for encouraging and useful exchanges, and the anonymous referee for a helpful report. The 2026 Debris Disk Connections conference at Cambridge University motivated further improvements.}  R.L. acknowledges support from the Heising-Simons Foundation 51 Pegasi b Fellowship. E.C. is grateful for a Simons Investigator grant and NSF grant 2205500 for ACCESS supercomputer allocations.

\software{Dustpy \citep{dustpy}, 
          Cupy \citep{cupy},
          Matplotlib \citep{Matplotlib}, 
          Numpy \& Scipy \citep{Numpy}.
          }




\bibliographystyle{aasjournal}
\bibliography{refs}




\end{document}